\documentstyle[twoside,fleqn,espcrc2,epsf]{article}

\title{
Two-flavour Schwinger model
with dynamical fermions
in the L\"uscher formalism 
}

\author{
S. Elser 
\thanks{supported by DFG research grant No.  WO 389/3-2; 
        \mbox{$\,$ email: elser@linde.physik.hu-berlin.de}}
and B. Bunk 
\thanks{email: bunk@linde.physik.hu-berlin.de} \\
{Institut f\"ur Physik, Humboldt--Universit\"at, Invalidenstr.110,
         10115 Berlin, Germany 
        }
       }
       
\begin{document}

\begin{abstract}
We report preliminary results for 2D massive QED with two flavours of Wilson
fermions, using the Hermitean variant of L\"uscher's bosonization
technique. The chiral condensate and meson masses are obtained.
The simplicity of the model allows for high statistics simulations close to
the chiral and continuum limit, both in the quenched approximation and with
dynamical fermions.
\end{abstract}

\maketitle

\section{Schwinger model}

Studies of algorithms for dynamical fermions are very time--consuming,
therefore we propose the massive Schwinger model (QED in two dimensions)
as a low-cost laboratory. We start from the Euclidean continuum Lagrangean
\begin{eqnarray}
{\cal L}
=
{1 \over 4} F^2
+
\sum_{a=1}^2 \Bigl[
\bar\psi^a ( / \hspace*{-0.195truecm} \partial 
+ i g / \hspace*{-0.25truecm} A + m)
\psi^a
\Bigr]
\nonumber
\end{eqnarray}
with two degenerate flavours of fermions.
For $m=0$ it is exactly soluble\cite{sch62}
and has a known expansion in the mass\cite{het95}.
Close to the chiral limit, we expect three light states (`pions'), massive
mesons and a non-zero fermion condensate.

The lattice version is defined using compact link variables $U_\mu(x) \in U(1)$
with the standard plaquette action
$S_g[U]$ and Wilson fermions with operator $M$.
Integration over the fermion variables results in
a positive effective action for the gauge fields:
\begin{eqnarray}
P_{\rm eff}[U] 
&\propto& 
{\rm det} M^2 \, e^{-S_g[U]} .
\nonumber
\end{eqnarray}

\section{L\"uscher's local bosonic theory}

As an alternative to the Hybrid Monte Carlo algorithm,
M. L\"uscher proposed a local bosonic formulation\cite{lue94}.
Let $Q=c\gamma_5 M = Q^\dagger$ be scaled so that its 
eigenvalues are in $[-1,1]$ and $P_{n}(s)$ a polynomial of even 
degree $n$ which approximates $1/s$ in $(0,1]$. 
Its roots $z_k$ ($k=1 \ldots n$) come in complex conjugate pairs
and determine $\sqrt{z_k}=\mu_k + i\nu_k$ ($\nu_k>0$).
Then
\begin{eqnarray}
P_{\rm eff}[U]
\!&\propto& \!\!
\det Q^2 \, e^{-S_g[U]}  
\nonumber \\
\!&=&\! \!
\det[Q^2 P_{n}(Q^2)] \, [\det P_{n}(Q^2)]^{-1} \, e^{-S_g[U]}     
\nonumber \\ 
\!&\propto&\!\!
C(Q^2) \, e^{-S_g[U]} \nonumber \\
&& \! \! \int \! 
{\cal D}\phi \, e^{-\sum_k [\phi_k^\dagger (Q-\mu_k)^2 \phi_k
+\nu^2_k \phi^\dagger_k \phi_k ]}
\nonumber
\end{eqnarray}
with $C(Q^2) \equiv \det[Q^2 P_{n}(Q^2)] \approx 1$
and $n$ complex bosonic Dirac fields $\phi_k$.

\section{Implementation}

We chose as approximation polynomials
$P_{n}(s)$ the Chebychev polynomials proposed by 
Bunk et al.\cite{bun95}. The convergence of
$P_{n}(s) \rightarrow 1/s$ as $n \rightarrow \infty$ is
exponential and uniform for $s \in [\epsilon,1]$,
introducing a parameter $\epsilon$ into the algorithm.

The updating process consists of
exact heat bath sweeps for the $\phi$'s and $U$'s\cite{bes79},
followed by a number of over-relaxation iterations.
We confirmed, in our preliminary runs,
that the reflection sweeps for the $\phi$'s and $U$'s have to be combined
in pairs, as was observed before\cite{jeg95}.
Finally, the approximation $C \approx 1$ has to be controlled. 
To this end we compute the lowest 8 eigenvalues\cite{ale95} of $Q^2$,
use them to estimate the change in $C$ and apply a global Metropolis
correction step with acceptance probability
$\min[1,{C' \over C}]$. This will be improved in the future.

In order to avoid trouble with bad pseudo-random numbers
we use L\"uscher's high-quality random 
number generator\cite{lue93}.

\section{Observables}

We measured averages of local fermion bilinears with a noisy estimator
scheme, using random spinors $\eta(x) = \pm 1$ as sources for the CG
solver. 
$\langle \bar\psi \gamma_5 \psi \rangle $
and 
$\langle \bar\psi \gamma_\mu \psi \rangle $
were checked to vanish within errors.
The fermion condensate $\langle \bar\psi \psi \rangle $ remains nonzero even
in the chiral limit because Wilson fermions break chiral invariance explicitly.
Our MC results are presented below.

For the determination of meson masses operators with various quantum
numbers are defined:
\begin{itemize}
\item
flavour triplet -
$( \bar\psi \gamma_5 \tau \psi )$ `$\pi$', 
$( \bar\psi \tau \psi )$ `$a_0$',
\item
flavour singlet -
$( \bar\psi \gamma_5 \psi )$ `$\eta$', 
$( \bar\psi \psi )$ `$f_0$' .
\end{itemize}
Moreover, insertion of a $\gamma_0$,
e.g. $( \bar\psi \gamma_0 \gamma_5 \tau \psi )$ for the $\pi$,
leads to alternative operators with
the same quantum numbers in the rest frame. This exhausts the Dirac algebra
in two dimensions.

The calculation of temporal correlators $\Gamma(\Delta t)$ involved
point sources at randomly chosen positions $(x,t)$ and summation over
spatial displacements $y$ in the other time slice $(y,t+\Delta t)$ to
project out the zero momentum states. As to the flavour--singlet channels,
the disconnected piece was subtracted with the aid of the noisy inversions
performed earlier for the measurement of the condensate.

\section{Calculations}

\subsection{Dynamical condensates}

The condensate with dynamical fermions was calculated
on 16x16 lattices for
$\beta = 2.0 $, $\kappa=0.20 \dots 0.275$
and
on 16x32 lattices for
$\beta = 10.0 $, $\kappa=0.20 \dots 0.25$.

\begin{figure}[tbp]
\epsfxsize=7.3cm
\epsffile{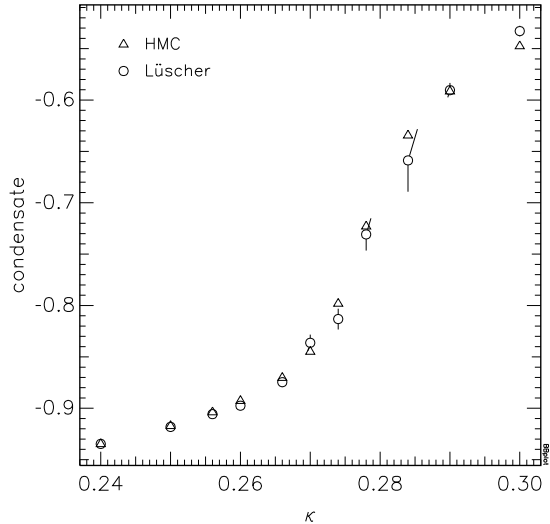}
\caption{Dynamical fermion condensate on a 16x16 lattice at $\beta=2.0$}
\label{dyncond}
\end{figure}

The L\"uscher algorithm required a small number ($n = 20 \ldots 40$)
of boson fields only and allowed for very high values of
$\epsilon = 0.01 \ldots 0.1$.
With moderate statistics ($200 \ldots 400$ measurements),
we reproduced the results obtained with a Hybrid Monte Carlo by
the Graz group\cite{lang96}, see Fig.~\ref{dyncond}.

\subsection{Quenched masses}

\begin{figure}[tbp]
\epsfxsize=7.3cm
\epsffile{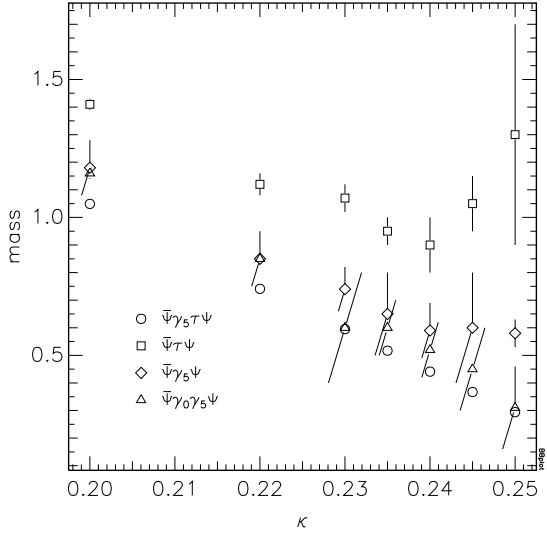}
\caption{Quenched meson masses on a 16x32 lattice at $\beta=6.0$}
\label{qu_masses}
\end{figure}

Meson masses in the quenched approximation were obtained on 16x32 lattices for
$\beta = 2.0 ,6.0 ,10.0, 20.0$ and $\kappa=0.20 \dots 0.275$
performing high-statistics runs with about
2000 independent measurements each. In case of $\beta = 6.0$,
Fig. \ref{qu_masses} shows the clear signal for the pion mass decreasing as
$\kappa$ is increased, its alternative operator gives masses which agree within
errors. For the $\eta$ and its variant, consistency is also verified with
larger errors. The higher states give noisy results.

\subsection{Dynamical masses}

With dynamical fermions, Irving et al.\cite{irv95} gave a mass of
$0.369(3)$ for a 32x32 lattice with $\beta=2.29$, $\kappa=0.26$.
We checked our implementation of the dynamical fermion update
with $n = 40$ and obtained $m = 0.377(4)$ from about 2000 measurements. 

\begin{figure}[tbp]
\epsfxsize=7.3cm
\epsffile{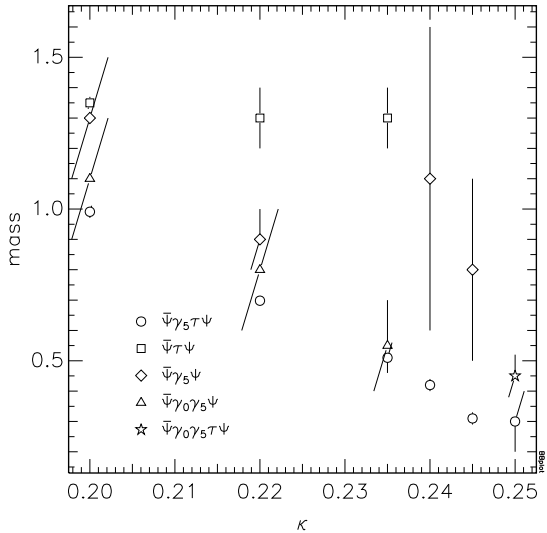}
\caption{Dynamical meson masses on a 16x32 lattice at $\beta=10.0$}
\label{dynmasses}
\end{figure}

More masses were determined on 16x32 lattices for
$\beta = 10.0 $, $\kappa=0.20 \dots 0.25$ and
are shown in Fig. \ref{dynmasses}.
We observe the same qualitative
behaviour as in the quenched case.

\section{Conclusions and outlook}

The two--flavour Schwinger
model has physical properties similar to QCD in four dimensions. It
is much easier to simulate even with dynamical fermions and 
allows to determine observables with high precision.
The preliminary results presented above are encouraging. 

The next steps will be to include standard versions of preconditioning,
a test of the non-hermitean variant of the polynomial approximation
and a noisy estimator which makes the Metropolis acceptance step exact.

As to the correlation functions, different noisy estimator schemes will
be tested and their efficiency compared.

Finally, the Schwinger model will be used to tune the boson algorithm and
to determine how the computational cost scales in the continuum limit
$a m_\eta \to 0$ with $m_\pi / m_\eta$ and $L m_\eta$ kept fixed. 
Although larger lattices than those used so far will be necessary, the
cost will be lower by orders of magnitude as compared to the case of
QCD in four dimensions. This makes the Schwinger model a reasonable
testing ground for dynamical fermion algorithms.

\section{Acknowledgements}
We thank the Graz group of Lang et al. for
their preliminary data to compare to
and U. Wolff, A.C. Irving and M. Peardon for discussions.

\end{document}